\documentclass[preprint,superscriptaddress,showpacs,preprintnumbers,amsmath,amssymb]{revtex4}

\usepackage{graphicx}
\usepackage{dcolumn}
\usepackage{bm}
\usepackage{amsmath,amssymb}

\def\lesssim{\ \raise.3ex\hbox{$<$}\kern-0.8em\lower.7ex\hbox{$\sim$}\ }
\def\gesim{\ \raise.3ex\hbox{$>$}\kern-0.8em\lower.7ex\hbox{$\sim$}\ }

\begin{document}
\title{Pseudogap temperature and effects of a harmonic trap in the BCS-BEC crossover regime of an ultracold Fermi gas}
\author{Shunji Tsuchiya}
\affiliation{Department of Physics, Tokyo University of Science, 1-3 Kagurazaka, Shinjuku-ku, Tokyo 162-8601, Japan}
\affiliation{Research and Education Center for Natural Sciences, Keio
University, 4-1-1 Hiyoshi, Kanagawa 223-8521, Japan}
\affiliation{CREST(JST), 4-1-8 Honcho, Saitama 332-0012, Japan}
\author{Ryota Watanabe}
\affiliation{Faculty of Science and Technology, Keio University, 3-14-1 Hiyoshi, Kohoku-ku, Yokohama 223-8522, Japan}
\author{Yoji Ohashi}%
\affiliation{Faculty of Science and Technology, Keio University, 3-14-1 Hiyoshi, Kohoku-ku, Yokohama 223-8522, Japan}
\affiliation{CREST(JST), 4-1-8 Honcho, Saitama 332-0012, Japan}

\date{\today}

\begin{abstract}
We theoretically investigate excitation properties in the pseudogap regime of a trapped Fermi gas. Using a combined $T$-matrix theory with the local density approximation, we calculate strong-coupling corrections to single-particle local density of states (LDOS), as well as the single-particle local spectral weight (LSW). Starting from the superfluid phase transition temperature $T_{\rm c}$, we clarify how the pseudogap structures in these quantities disappear with increasing the temperature. As in the case of a uniform Fermi gas, LDOS and LSW give different pseudogap temperatures $T^*$ and $T^{**}$ at which the pseudogap structures in these quantities completely disappear. Determining $T^*$ and $T^{**}$ over the entire BCS (Bardeen-Cooper-Schrieffer)-BEC (Bose-Einstein condensate) crossover region, we identify the pseudogap regime in the phase diagram with respect to the temperature and the interaction strength. We also show that the so-called back-bending peak recently observed in the photoemission spectra by JILA group may be explained as an effect of pseudogap phenomenon in the trap center. Since strong pairing fluctuations, spatial inhomogeneity, and finite temperatures, are important keys in considering real cold Fermi gases, our results would be useful for clarifying normal state properties of this strongly interacting Fermi system.
\end{abstract}
\maketitle

\section{Introduction}\label{section1}

The pseudogap temperature is a fundamental quantity in strongly
interacting Fermi superfluids, such as high-$T_{\rm c}$
cuprates\cite{Damascelli,Fischer,Lee} and superfluid $^{40}$K and $^{6}$Li Fermi
gases\cite{reviews,Regal,Zwierlein,Kinast,Bartenstein,Eagles,Leggett,NSR,SadeMelo,Timmermans,Holland,Ohashi,Ohashi2,Ohashi3}. The
normal state region below this characteristic temperature is referred to
as the pseudogap regime, where anomalies are seen in various physical
quantities, such as a gap-like structure in single-particle density of
states. As the origin of the pseudogap regime, the so-called
preformed-pair scenario has been extensively discussed in high-$T_{\rm
c}$ cuprates\cite{Randeria,Singer,Janko,Yanase,Rohe,Perali}. However, because of the complexity of this
system, other scenarios have been also proposed, such as
antiferromagnetic spin fluctuations\cite{Pines,Kampf}, and a hidden
order\cite{Chakravarty}. In contrast, in cold Fermi gases, because of
the simplicity of the system, the origin can be uniquely identified as
strong pairing
fluctuations\cite{Stewart,Gaebler,Tsuchiya,Tsuchiya2,Watanabe,Chen,Magierski,Hu,Perali2,Su,Mueller}. Thus,
this system would be very useful for the study of the preformed-pair
scenario. 
\par
In considering the pseudogap phenomenon, we note that this is a normal state phenomenon, free from any phase transition at the pseudogap temperature. Because of this, the pseudogap temperature may depend on what we measure. In a previous paper\cite{Tsuchiya} for a uniform Fermi gas, we showed that the pseudogap temperature $T^*$ which is defined as the temperature at which the pseudogap structure disappears in the single-particle density of states is different from the pseudogap temperature $T^{**}$ determined from the single-particle spectral weight. While one finds $T^*>T^{**}$ in the weak-coupling BCS regime, $T^{**}$ becomes higher than $T^*$, as one passes through the BCS-BEC crossover region. 
\par
Since a cold Fermi gas is always trapped in a harmonic potential, spatial inhomogeneity is also a key in considering the pseudogap problem of a real Fermi gas. Indeed, in the crossover region at $T_{\rm c}$, it has been shown \cite{Tsuchiya2} that, while a clear pseudogap structure can be seen in the single-particle excitation spectrum in the trap center, one only sees a free-particle-like dispersion around the edge of the gas. Since the recent photoemission-type experiment developed by JILA group does not have spatial resolution, the observed spectra correspond to the sum of spatially inhomogeneous excitation spectra. Including this, we showed\cite{Tsuchiya2} that the anomalous photoemission spectra observed at $T_{\rm c}$ can be theoretically reproduced. We briefly note that the importance of such inhomogeneous pseudogap phenomena in cold Fermi gases has been also pointed out in Ref.\cite{Hu}.
\par
In this paper, we theoretically identify the pseudogap regime of a
trapped Fermi gas in the BCS-BEC crossover region. Extending our
previous work at $T_{\rm c}$\cite{Tsuchiya2} to the region above $T_{\rm
c}$, we calculate the single-particle local density of states (LDOS), as
well as the single-particle local spectral weight (LSW), within the
framework of a combined strong-coupling $T$-matrix theory with the local
density approximation (LDA). While the LDOS is suitable for the understanding of the meaning of `pseudo'-gap, the LSW is closer to the photoemission spectrum. Starting from $T_{\rm c}$, we show how the pseudogap structures in these quantities gradually disappear, as one increases the temperature. We define the pseudogap temperature $T^*$ as the temperature at which the pseudogap structure completely disappears in LDOS. In the same manner, we also define another pseudogap temperature $T^{**}$ for LSW. We determine both $T^*$ and $T^{**}$ in the entire BCS-BEC crossover region. Using these two pseudogap temperatures, we identify the pseudogap region in the phase diagram of cold Fermi gases. We also examine the photoemission spectrum, and explain the so-called back-bending behavior observed in $^{40}$K Fermi gases from the viewpoint of inhomogeneous pseudogap effect.
\par
This paper is organized as follow. In Sec.~\ref{section2}, we explain our formulation. Using the combined $T$-matrix theory with LDA, we first determine the chemical potential above $T_{\rm c}$ in the BCS-BEC crossover. We then calculate LDOS, as well as LSW, above $T_{\rm c}$. Here, we also calculate the photoemission spectrum for a trapped Fermi gas within LDA. In Sec.~\ref{section3}, we examine pseudogap effects on the LDOS and LSW, focusing on their temperature dependences. After determining the pseudogap temperatures $T^*$ and $T^{**}$, we identify the pseudogap region in the phase diagram of trapped Fermi gases above $T_{\rm c}$. In Sec. IV, we discuss the photoemission spectrum. We clarify how the pseudogap affects this quantity. Throughout this paper, we take $\hbar=k_{\rm B}=1$.

\section{Formulation}
\label{section2}
We consider a two-component Fermi gas, described by the BCS Hamiltonian,
\begin{equation}
H-\mu N=\sum_{{\bm p},\sigma}\xi_{\bm p} c_{\bm p\sigma}^\dagger c_{\bm p\sigma}
-U\sum_{\bm p,\bm p',\bm q}c_{{\bm p}+{\bm q}/2\uparrow}^\dagger
 c_{-{\bm p}+{\bm q}/2\downarrow}^\dagger c_{-{\bm p}'+{\bm
 q}/2\downarrow}c_{{\bm p}'+{\bm q}/2\uparrow},
\label{ham0}
\end{equation}
where $c^\dagger_{\bm p\sigma}$ is the creation operator of a Fermi atom
with pseudospin $\sigma=\uparrow,\downarrow$, which describe two atomic
hyperfine states. $\xi_{\bm p}=\varepsilon_{\bm p}-\mu=\frac{{\bm
p}^2}{2m}-\mu$ is the kinetic energy, measured from the chemical
potential $\mu$ (where $m$ is an atomic mass). Here, $N$ is the total
number operator of Fermi atoms. $-U(<0)$ is an attractive
interaction which can be tuned by a Feshbach resonance\cite{Chin}.
$U$ is related to the $s$-wave scattering length $a_s$ as
$4\pi a_s/m=-U/[1-U\sum_{\bm p}^{\omega_c}1/(2\epsilon_p)]$ (where
$\omega_c$ is a high-energy cutoff)\cite{Randeria2}. As usual, we conveniently measure
the interaction strength in terms of the inverse scattering length
$(k_Fa_s)^{-1}$, where $k_{\rm F}$ is the Fermi momentum. We will
include effects of a trap later.  
\par
\begin{figure}
\centerline{\includegraphics[width=12cm]{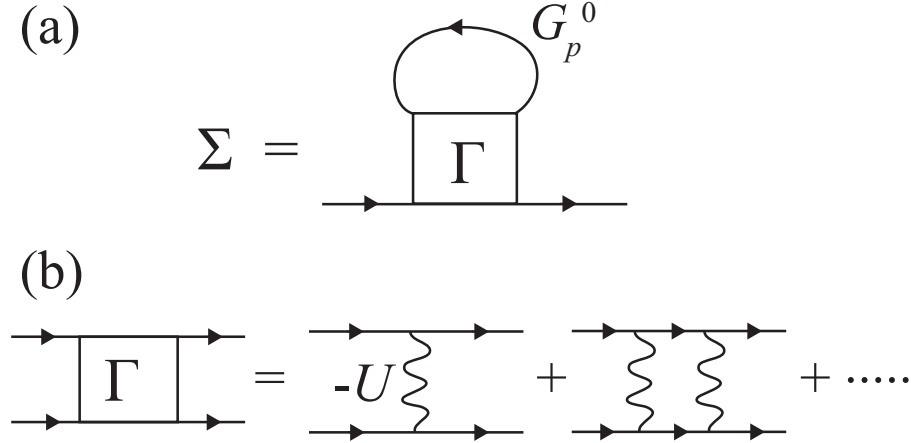}}
\caption{(a) Self-energy $\Sigma_{\bm p}(i\omega_n)$, and (b) particle-particle scattering matrix $\Gamma_{\bm q}(i\nu_n)$, in the $T$-matrix approximation. The solid and wavy lines represent the noninteracting Green's function $G_{\bm p}^0(i\omega_n)$ and the pairing interaction $-U$, respectively.} 
\label{fig1}
\end{figure}

We treat pairing fluctuations within the standard $T$-matrix
theory\cite{Janko,Rohe,Perali,Tsuchiya,Tsuchiya2,Watanabe}. Despite its simplicity, previous studies have shown that this strong-coupling theory gives quantitatively reliable results for pseudogap phenomena\cite{Tsuchiya2}. Within this framework, the single-particle thermal Green's function is given by
\begin{equation}
G_{\bm p}(i\omega_n)=\frac{1}{G^0_{\bm p}(i\omega_n)^{-1}-\Sigma_{\bm p}(i\omega_n)}, 
\label{Green}
\end{equation}
where $\omega_n$ is the fermion Matsubara frequency. $G^0_{\bm
p}(i\omega_n)= 1/[i\omega_n-\xi_{\bm p}]$ is the single-particle Green's
function for a free fermion. The self-energy part $\Sigma_{\bm
p}(i\omega_n)$ involves effects of pairing fluctuations within the
$T$-matrix approximation, which is diagrammatically given in
Fig.~\ref{fig1}. Summing up these diagrams, we obtain 
\begin{equation}
\Sigma_{\bm p}(i\omega_n)=T\sum_{\bm q,\nu_n}
\Gamma_{\bm q}(i\nu_n)G^0_{\bm q-\bm p}(i\nu_n-i\omega_n),
\label{eq.self}
\end{equation}
where $\nu_n$ is the boson Matsubara frequency. The particle-particle scattering matrix $\Gamma_{\bm q}(i\nu_n)$ is given by
\begin{eqnarray}
\Gamma_{\bm q}(i\nu_n)&=&\frac{-U}{1-U\Pi_{\bm q}(i\nu_n)}\nonumber\\
&=&\frac{4\pi a_s}{m}\frac{1}{1+\frac{4\pi a_s}{m}
\left[\Pi_{\bm
q}(i\nu_n)-\sum_{\bm p}\frac{1}{2\varepsilon_{\bm p}}
\right]}.                                                   
\label{eq.gamma}
\end{eqnarray}
The pair-correlation function 
\begin{eqnarray}
\Pi_{\bm q}(i\nu_n)
&=&T\sum_{\bm p,\omega_n}
G_{\bm p+\bm q/2}^0(i\nu_n+i\omega_n)G^0_{-\bm p+\bm q/2}(-i\omega_n)
\nonumber
\\
&=&
\sum_{\bm q}
{
1-f(\xi_{\bm p+\bm q/2})-f(\xi_{-\bm p+\bm q/2})
\over
\xi_{\bm p+\bm q/2}+\xi_{-\bm p+\bm q/2}-i\nu_n
},
\label{eq.pi}
\end{eqnarray}
describes fluctuations in the Cooper channel, where $f(\xi)$ is the Fermi distribution function.
\par
We now include effects of a harmonic trap $V(r)=m\omega_{\rm tr}^2r^2/2$ within LDA (where $\omega_{\rm tr}$ is a trap frequency). This extension is simply achieved by replacing the Fermi chemical potential $\mu$ by the LDA expression $\mu(r)\equiv \mu-V(r)$\cite{Pethick} in Eqs. (\ref{Green})-(\ref{eq.pi}). In this paper, we explicitly write the variable $r$ for LDA quantities. For example, the LDA single-particle Green's function is written as 
\begin{equation}
G_{\bm p}(i\omega_n,r)=\frac{1}
{i\omega_n-\xi_{\bm p}(r)-\Sigma_{\bm p}(i\omega_n,r)},
\label{Green_LDA}
\end{equation}
where $\xi_{\bm p}(r)=\varepsilon_{\bm p}-\mu(r)$. 
\par
To examine the pseudogap phenomenon in a trapped Fermi gas, we consider the single-particle local density of states (LDOS) $\rho(\omega,r)$, as well as the single-particle local spectral weight (LSW) $A({\bm
p},\omega,r)$. Their LDA expressions are given by
\begin{eqnarray}
\rho(\omega,r)=-\frac{1}{\pi}\sum_{\bm p}
{\rm Im}[G_{\bm p}(i\omega_n\to\omega+i\delta,r)],
\label{eq.ldos}
\end{eqnarray}
\begin{eqnarray}
A(\bm p,\omega,r)=-\frac{1}{\pi}
{\rm Im}[G_{\bm p}(i\omega_n\to\omega+i\delta,r)].
\label{eq.lsw}
\end{eqnarray}
The former can be written as $\rho(\omega,r)=\sum_{\bm p}A({\bm p},\omega,r)$, so that LSW may be viewed as the momentum resolved LDOS.
\par
In this paper, we also consider the photoemission-type experiment developed by JILA group\cite{Stewart,Gaebler}. When the $\uparrow$-spin state is coupled with another hyperfine state $|3\rangle$ ($\ne |\uparrow\rangle,|\downarrow\rangle$) by radio-frequency (rf) pulse\cite{Stewart,Gaebler}, the photoemission spectrum is given by
\begin{equation}
I_{\rm ave}(\bm p,\Omega)=\frac{2\pi t_F^2}{V}\int d{\bm r}
A(\bm p,\xi_{\bm p}(r)-\Omega,r)f(\xi_{\bm p}(r)-\Omega).
\label{eq.photo}
\end{equation}
(We summarize the derivation of Eq. (\ref{eq.photo}) in the Appendix.) Here, $V=4\pi R_F^3/3$ (where $R_F=\sqrt{2\mu/(m\omega_{\rm tr}^2)}$ is the Thomas-Fermi radius\cite{Pethick}), and $t_F$ is a coupling between the $\uparrow$-spin state and $|3\rangle$. Since the current experiment has no spatial resolution, we have taken the spatial average in Eq. (\ref{eq.photo}). Apart from this spatial integration, the photoemission spectrum is closely related to the spectral weight $A({\bm p},\omega,r)$. 
\par
To directly compare our results with the experimental data\cite{Stewart,Gaebler}, we slightly modify Eq. (\ref{eq.photo}) as
\begin{eqnarray}
\overline{A(\bm p,\omega)f(\omega)}&\equiv& I_{\rm ave}(\bm
 p,\Omega\to\xi_{\bm p}-\omega)\nonumber\\
&=&\frac{2\pi t_{\rm F}^2}{V}\int d{\bm r}A(\bm
 p,\omega-\mu(\bm r),r) f(\omega-\mu(\bm r)).
\label{eq.sw}
\end{eqnarray}
For a free Fermi gas, Eq. (\ref{eq.sw}) is evaluated to give
\begin{eqnarray}
\overline{A(\bm p,\omega)f(\omega)}=-\frac{3\pi^{3/2}t_{\rm
 F}^2}{2(\beta\mu)^{3/2}}{\rm Li}_{3/2}(-e^{-\beta\omega})\delta(\omega-\xi_{\bm
 p}),
\label{eq.freeswT}
\end{eqnarray}
where ${\rm Li}_s(z)$ is the polylogarithm of $z$ in the order $s$.
The peak energy of Eq. (\ref{eq.freeswT}) gives the single-particle
dispersion $\omega=\xi_{\bm p}$.
At $T=0$, Eq. (\ref{eq.sw}) reduces to
\begin{equation}
\overline{A({\bm p},\omega)f(\omega)}=2\pi t_{\rm
 F}^2\left|\frac{\omega}{\mu}\right|^{3/2}\delta(\omega-\xi_{\bm p})\theta(-\omega).
\label{eq.freesw}
\end{equation}
\par
JILA's experiments\cite{Stewart,Gaebler} have also examined the occupied density of states, defined by
\begin{equation}
\overline{\rho(\omega)f(\omega)}\equiv
{1 \over 2\pi t_{\rm F}^2}
\sum_{\bm p}I_{\rm ave}(\bm p,\Omega\to\xi_{\bm p}-\omega).
\label{eq.dos}
\end{equation}
For a free Fermi gas, Eq. (\ref{eq.dos}) gives
\begin{equation}
\overline{\rho(\omega)f(\omega)}=-\frac{3m^{3/2}}{4\sqrt{2}\pi^{3/2}}\frac{1}{(\beta\mu)^{
3/2}}{\rm
Li}_{3/2}(-e^{-\beta\omega})\sqrt{\omega+\mu}\theta(\omega+\mu).
\label{eq.dosT}
\end{equation}
At $T=0$, Eq. (\ref{eq.dosT}) reduces to
\begin{equation}
\overline{\rho(\omega)f(\omega)}=\frac{m^{3/2}}{\sqrt{2}\pi^2}\left|\frac{\omega}{\mu}\right|^{3/2}\sqrt{\omega+\mu}\theta(\omega+\mu)\theta(-\omega).
\label{eq.freedos}
\end{equation}
Equation (\ref{eq.freedos}) is just the ordinary density of states $\propto\sqrt{\omega+\mu}$ multiplied by $|\omega|^{3/2}$ when $\omega\le 0$ and $\omega+\mu\ge 0$.
\par
\begin{figure}
\centerline{\includegraphics[width=12cm]{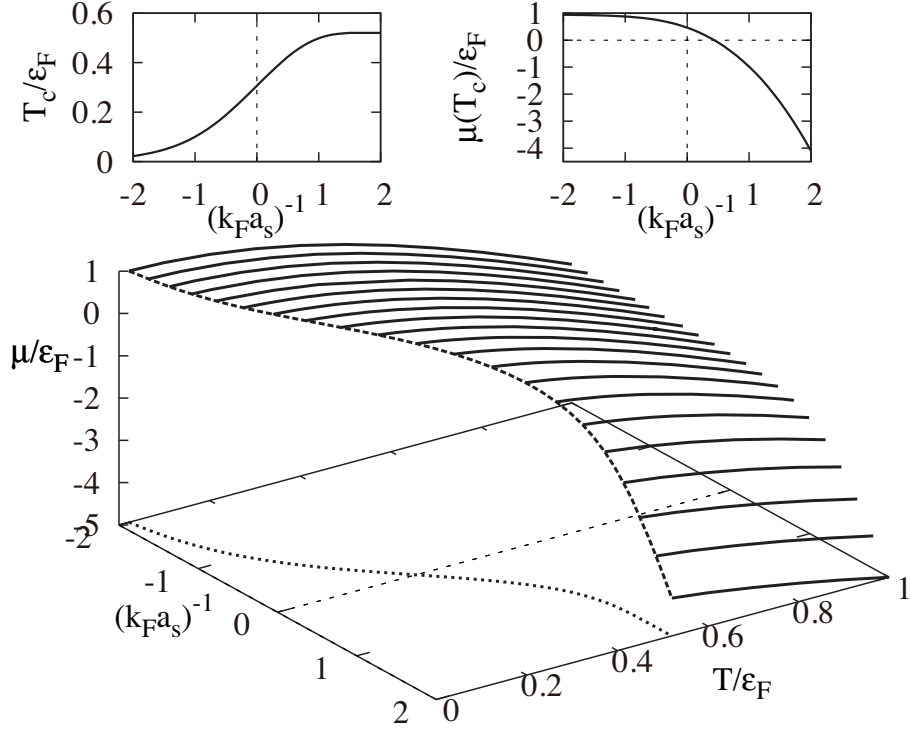}}
\caption{Calculated Fermi chemical potential $\mu$ above $T_{\rm
 c}$. $\varepsilon_{\rm F}=(3N)^{1/3}\omega_{\rm tr}$ is the Fermi
 energy of a trapped Fermi gas, and the interaction is measured in terms
 of the inverse scattering length $(k_{\rm F}a_s)^{-1}$, where $k_{\rm
 F}$ is the Fermi momentum. We will use this result in calculating
 Eqs.(\ref{eq.ldos})-(\ref{eq.photo}) in Secs. III and IV. The upper left and right panels show $T_{\rm c}$ and $\mu(T_{\rm c})$, respectively.} 
\label{fig2}
\end{figure}
In order to calculate Eqs. (\ref{eq.ldos}), (\ref{eq.lsw}),
(\ref{eq.sw}), and (\ref{eq.dos}), we need to determine the Fermi
chemical potential $\mu$ from the equation for the total number $N$ of
Fermi atoms. The LDA number equation is given by 
\begin{equation}
N=2T\int d\bm r\sum_{\bm p,\omega_n}G_{\bm p}(i\omega_n,r)e^{i\omega_n\delta}.
\label{number}
\end{equation}
The calculated chemical potential $\mu(T\ge T_{\rm c})$ is shown in Fig. \ref{fig2}. We briefly note that the LDA superfluid phase transition temperature $T_{\rm c}$ is given as the temperature at which the Thouless criterion\cite{Thouless} is satisfied in the trap center ($r=0$)\cite{Ohashi2}. The resulting LDA $T_{\rm c}$ equation is given by
\begin{equation}
\Gamma_{\bm q=0}(i\nu_n=0,r=0)^{-1}=0.
\label{Thouless}
\end{equation}
\par
\begin{figure}
\centerline{\includegraphics[width=12cm]{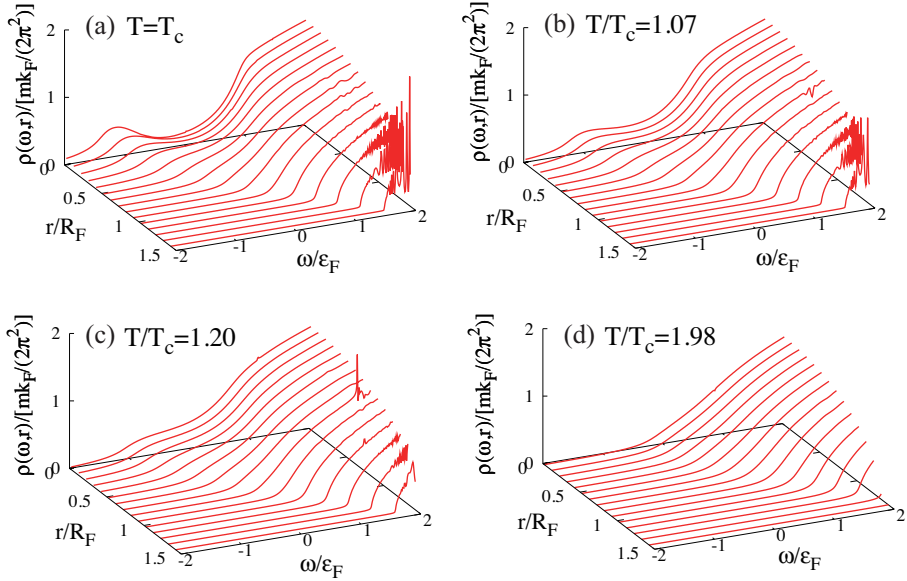}}
\caption{(Color online) Calculated local density of states (LDOS)
 $\rho(\omega,r)$ above $T_{\rm c}$ in the unitarity limit
 ($(k_Fa_s)^{-1}=0$). The fine peaks  seen in LDOS at large $r$ are a
 computational problem in the momentum integration of $A({\bm
 p},\omega,r)$, involving $\delta$-functional peaks as $A({\bm
 p},\omega,r)\simeq \delta(\omega-\xi_{\rm p}(r))$.	
}
\label{fig3}
\end{figure}

\section{pseudogap temperatures determined from LDOS and LSW}
\label{section3}
Figure \ref{fig3} shows the LDOS $\rho(\omega,r)$ in the unitarity limit ($(k_{\rm F}a_s)^{-1}=0$). At $T_{\rm c}$, panel (a) shows that LDOS in the trap center ($r=0$) has a large dip structure around $\omega=0$. Since the superfluid order parameter vanishes at $T_{\rm c}$, this pseudogap structure purely arises from strong pairing fluctuations in the unitarity regime\cite{Tsuchiya,Watanabe,Tsuchiya2}. The pseudogap (dip) structure gradually disappears, as one goes away from the trap center, because pairing fluctuations become weak around the edge of the gas ($r\simeq R_F$) due to the low particle density. 
\par
With regard to inhomogeneous pairing fluctuations, we note that pairing fluctuations are described by the analytic continued particle-particle scattering matrix, 
\begin{equation}
\Gamma_{\bm q}(i\nu_n\to\omega+i\delta,r)=
\frac{4\pi a_s}{m}\frac{1}{1+\frac{4\pi a_s}{m}
\left[\Pi_{\bm
q}(i\nu_n\to\omega+i\delta,r)-\sum_{\bm p}\frac{1}{2\varepsilon_{\bm p}}
\right]},
\label{an-gamma}
\end{equation}
where
\begin{eqnarray}
\Pi_{\bm q}(i\nu_n\to\omega+i\delta,r)
=
\sum_{\bm q}
{
1-f(\xi_{\bm p+\bm q/2}(r))-f(\xi_{-\bm p+\bm q/2}(r))
\over
\xi_{\bm p+\bm q/2}(r)+\xi_{-\bm p+\bm q/2}(r)-(\omega+i\delta)
}.
\label{an-pi}
\end{eqnarray}
Because of the Thouless criterion in Eq. (\ref{Thouless}), in the trap center ($r=0$), Eq. (\ref{an-gamma}) with $\omega={\bm q}=0$ diverges at $T_{\rm c}$. On the other hand, because the LDA chemical potential $\mu(r)=\mu-V(r)=\mu-m\omega^2r^2/2$ decreases as one goes away from the trap center ($r>0$), $\Pi_{{\bm q}=0}(0,r\ne 0)$ becomes small. Thus, low energy and low momentum pairing fluctuations described by $\Gamma_{{\bm q}\simeq 0}(\omega\simeq 0,r>0)$ also become weaker for larger $r$. In particular, around the edge of the gas cloud, panel (a) shows that the LDOS becomes close to that of a noninteracting Fermi gas,
\begin{equation}
\rho(\omega,r)={m^{3\over 2} \over \sqrt{2}\pi^2}\sqrt{\omega+\mu(r)}.
\label{eq.fdos}
\end{equation}
\par
As one increases the temperature, the pseudogap structure in LDOS gradually disappears from the outer region of the gas cloud, as shown in Figs. \ref{fig3}(b)-(d). Defining the pseudogap temperature $T^*$ as the temperature at which the dip structure in the trap center disappears, one obtains $T^*=1.1T_{\rm c}$ in the case of Fig. \ref{fig3}. Above the pseudogap temperature $T^*$, Fig.\ref{fig3}(d) shows that the overall structure of LDOS becomes close to that for a free Fermi gas in Eq. (\ref{eq.fdos}), although, in the trap center, LDOS around the threshold energy ($\omega=-\mu(r)$) is found to be still affected by pairing fluctuations to deviate from $\rho(\omega,r)\sim\sqrt{\omega+\mu(r)}$.
\par
\begin{figure}
\centerline{\includegraphics[width=12cm]{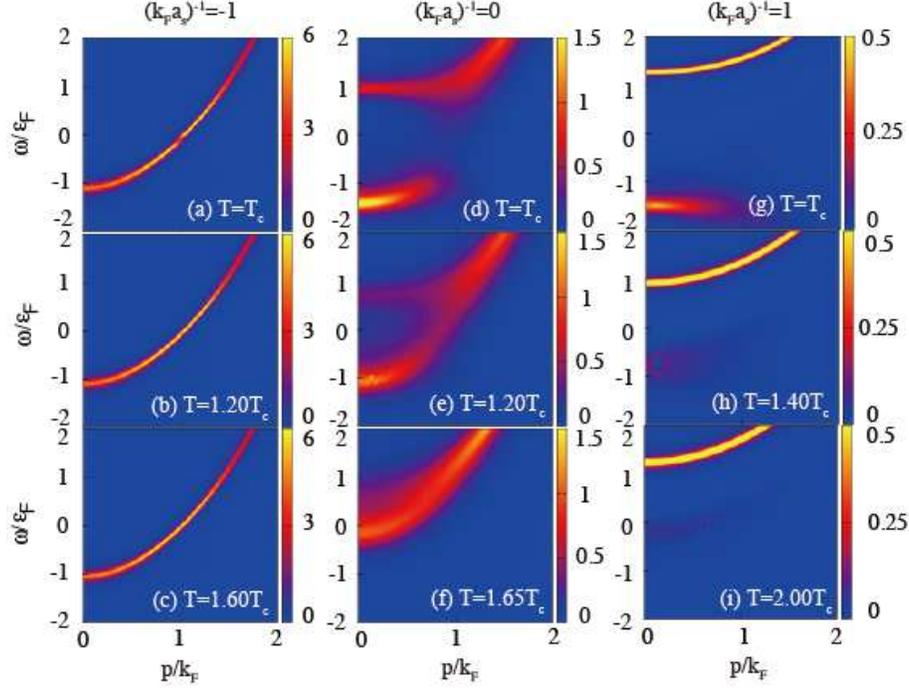}}
\caption{(Color online) Calculated intensity of local spectral weight (LSW) $A(\bm p,\omega)$ in the trap center $r=0$. The intensity is normalized by the inverse Fermi energy $\varepsilon_F^{-1}$.}
\label{fig4}
\end{figure}

Figure \ref{fig4} shows the LSW $A(\bm p,\omega,r)$ at $r=0$. In the BCS
side (panels (a)-(c)), although a pseudogap structure can be slightly
seen at $p\simeq k_{\rm F}$ in panel (a), it soon disappears at higher
temperatures. (See panels (b) and (c).) As discussed in
Ref. \cite{Tsuchiya2}, the pseudogap effects on LSW soon disappears, as
one moves away from the trap center (although we do not explicitly show the
spatial dependence of LSW in the BCS side in this paper). Thus,
Fig. \ref{fig4}(a)-(c) indicate that LSW is not useful for the
observation of pseudogap phenomenon in the BCS side. 
\par
In the unitarity limit, we see a clear pseudogap structure in LSW at $r=0$,  as shown in Figs. \ref{fig4}(d)-(f). In particular, at $T_{\rm c}$, although the superfluid order parameter vanishes, the overall pseudogap structure is similar to the superfluid gap structure in the BCS spectral weight, given by, 
\begin{equation}
A(\bm p,\omega,r=0)=u_{\bm p}^2\delta(\omega-E_{\bm p})+v_{\bm p}^2\delta(\omega+E_{\bm p}),
\label{BCSSW}
\end{equation}
where $E_{\bm p}=\sqrt{\xi_{\bm p}^2+\Delta^2}$, $u_{\bm p}=\sqrt{(1+\xi_{\bm p}/E_{\bm p})/2}$, and $v_{\bm p}=\sqrt{(1-\xi_{\bm p}/E_{\bm p})/2}$. Here, $\Delta$ is the BCS superfluid order parameter. In the BCS case, the double peak structure at $\omega=\pm E_{\bm p}$ may be regarded as a result of a coupling between particle and hole excitations by the superfluid order parameter $\Delta$. Indeed, the diagonal component of the BCS Green's function can be written in the form
\begin{eqnarray}
G_{11}({\bm p},i\omega_n)
&=&
-{i\omega_n+\xi_{\bm p} \over \omega_n^2+\xi_{\bm p}^2+\Delta^2}
\nonumber
\\
&=&
{1 \over \displaystyle (i\omega_n-\xi_{\bm p})-{\Delta^2 \over i\omega_n+\xi_{\bm p}}}.
\label{bcsgreen}
\end{eqnarray}
Noting that $1/(i\omega_n-\xi_{\bm p})$ and $1/(i\omega_n+\xi_{\bm p})$ represent the particle and hole Green's functions, respectively, $\Delta^2$ in Eq. (\ref{bcsgreen}) is found to work as a coupling between the particle branch and the hole branch.  In the pseudogap case at $T_{\rm c}$, since the particle-particle scattering matrix $\Gamma_{\bm q}(i\nu_n)$ diverges at $\nu_n={\bm q}=0$ (Thouless criterion), one may approximate the self-energy in Eq. (\ref{eq.self}) to 
\begin{equation}
\Sigma_{\bm p}(i\omega_n)\simeq T\sum_{{\bm q},\nu_n}
\Gamma_{\bm q}(i\nu_n)\times G^0_{-{\bm p}}(-i\omega_n).
\label{eq.self2}
\end{equation}
When one substitutes Eq. (\ref{eq.self2}) into Eq. (\ref{Green}), the resulting expression has the same form as Eq. (\ref{bcsgreen}) where $\Delta^2$ is replaced by the {\it pseudogap parameter} $\Delta_{\rm pg}^2\equiv -T\sum_{{\bm q},\nu_n}\Gamma_{\bm q}(i\nu_n)$\cite{Perali,Tsuchiya}. That is, pairing fluctuations described by $\Gamma_{\bm q}(i\nu_n)$ induces a particle-hole coupling, leading to the pseudogap structure seen in Fig. \ref{fig4}(d). Since pairing fluctuations become weak at higher temperatures, the double peak structure in the spectrum becomes obscure to eventually disappear, as shown in Figs. \ref{fig4}(e) and (f). 
\par
Besides the particle-hole coupling, pairing fluctuations also lead to finite lifetime of quasiparticle excitations. Because of this, the LSW in the unitarity limit exhibits broader spectra compared with that in the BCS regime, as shown in Figs. \ref{fig4} (a)-(f).
\par
In the strong-coupling BEC limit, the system reduces to a gas of
two-body bound molecules, so that single-particle excitations are simply
described by dissociations of these two-body bound states. Then, noting
that the single-particle spectral weight in the negative energy region
($\omega<0$) physically describes hole excitations, we expect that the
intensity of the lower spectral branch becomes weak in this {\it
two-body} regime. Indeed, the lower peak is found to be very weak in
Fig. \ref{fig4}(g)-(i). We briefly note that the peak width of the upper
branch in the BEC regime is sharper than the case of unitarity limit
shown in Figs. \ref{fig4}(d)-(f), which is due to the fact that this
branch simply describes the dissociation energy of a two-body bound
state in the BEC limit.
\par
\begin{figure}
\centerline{\includegraphics[width=8cm]{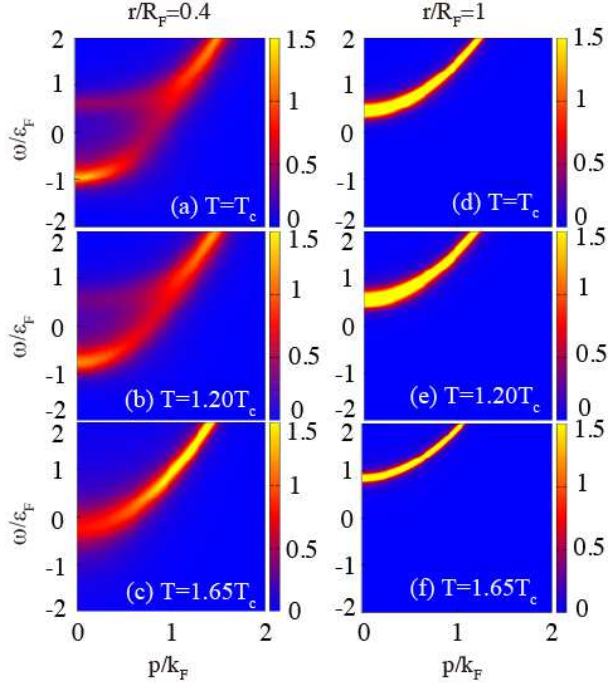}}
\caption{(Color online) Calculated intensity of LSW $A(\bm p,\omega,r)$
 in the unitarity limit ($(k_Fa_s)^{-1}=0$). The intensity is normalized by the inverse Fermi energy $\varepsilon_F^{-1}$.}
\label{swspat_asi0}
\end{figure}
\begin{figure}
\centerline{\includegraphics[width=8cm]{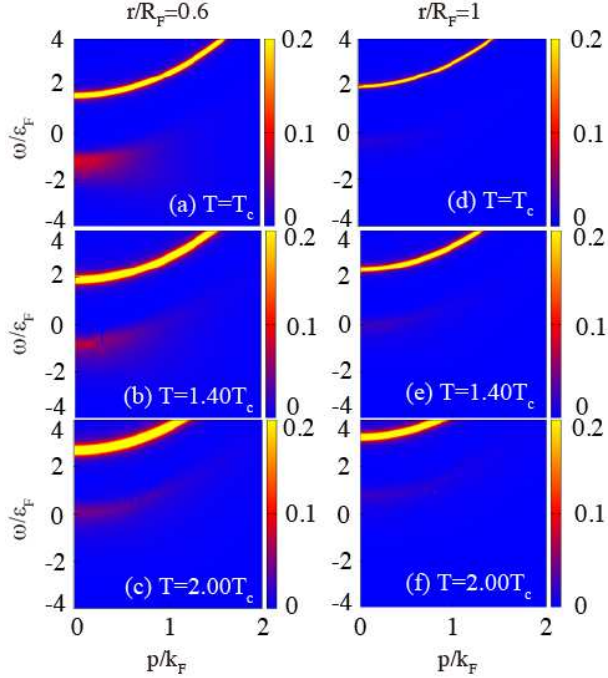}}
\caption{(Color online) Calculated intensity of LSW $A(\bm p,\omega,r)$
 in the BEC side ($(k_Fa_s)^{-1}=1$). The intensity is normalized by the inverse Fermi energy $\varepsilon_F^{-1}$.}
\label{swspat_asi10}
\end{figure}
Figures \ref{swspat_asi0} and \ref{swspat_asi10}, respectively, show the LSW in the unitarity limit ($(k_Fa_s)^{-1}=0$) and in the BEC regime ($(k_Fa_s)^{-1}=1$). We find that the pseudogap effect is not so remarkable around the edge of the gas cloud. We also find that the $r$-dependence of LSW in Fig. \ref{swspat_asi0} is similar to the temperature dependence of this quantity shown in Figs. \ref{fig4} (d)-(f). To understand this similarity, we recall that the spatial dependence of LSW is dominated by the LDA chemical potential $\mu(r)=\mu-V(r)$, which decreases as $r$ increases. The chemical potential $\mu$ also becomes small with increasing the temperature, as seen in Fig. \ref{fig2}. As a result, the increase of $r$ and the increase of the temperature lead to the similar effect on LSW seen in Figs. \ref{fig4}(d)-(f) and Fig. \ref{swspat_asi0}\cite{notezz}. 
\par
The lower peak in the BCS spectral weight in Eq. (\ref{BCSSW}) at $\omega=-E_{\bm p}$ is a downward curve in the BEC regime where $\mu<0$. Such a behavior is also seen in the lower peak line, as shown in Fig. \ref{fig4}(g). However, Figs. \ref{fig4}(h) and (i) show that this downward dispersion gradually changes into an upward one with increasing the temperature. The resulting upward dispersion of the lower branch is similar to the hole branch in the presence of a Fermi surface ($\mu>0$), so that this phenomenon would have to do with the hole-type character of the lower branch. At high temperatures, since unpaired fermions are thermally excited to occupy the upper branch of the excitation spectrum, these free-particle-like atoms are expected to give the upward spectrum in LSW. 
\par
\begin{figure}
\centerline{\includegraphics[width=10cm]{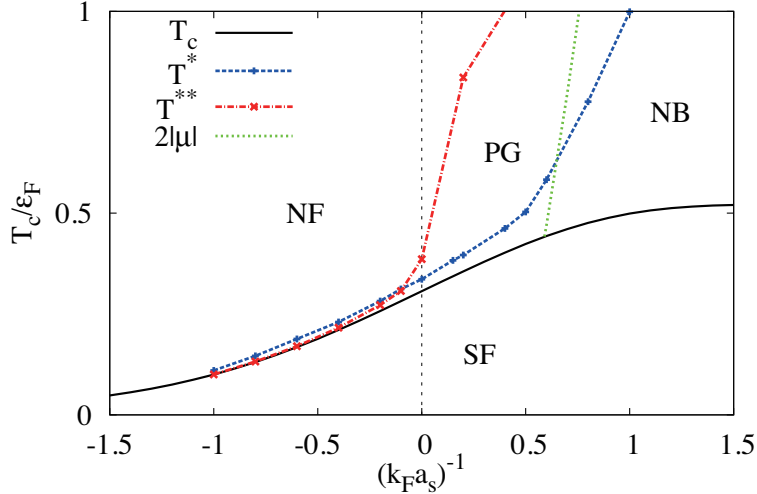}}
\caption{(Color online) Two pseudogap temperatures $T^*$ and $T^{**}$ determined from LDOS and LSW, respectively. The solid line shows the superfluid transition temperature $T_{\rm c}$, below which the system is in the superfluid phase (SF). In this figure, we also plot $2|\mu(T_{\rm c})|$ in the normal state of the BEC regime where $\mu(T_{\rm c})<0$. Since $2|\mu|$ equals the dissociation energy of a two-body bound state in the BEC limit, $T\sim 2|\mu(T_{\rm c})|$ physically gives the characteristic temperature where two-body bound molecules are formed. Thus, the right side of $2|\mu(T_{\rm c})|$ may be viewed as a normal-state molecular Bose gas (NB), rather than a Fermi gas. The pseudogap regime (PG) is the region surrounded by $T_{\rm c}$, $2|\mu|$, and $T^*$ or $T^{**}$. The left side of the pseudogap regime is the normal Fermi gas regime (NF), where strong-coupling effects are not crucial. We note that $T^*$, $T^{**}$, and $2\mu(T_{\rm c})$, are all characteristic temperatures, free from any phase transition.}
\label{fig5}
\end{figure}
\par
As in the unitarity limit, the $r$-dependence of LSW in Fig. \ref{swspat_asi10} is
similar to the temperature dependence of this quantity in Figs. \ref{fig4}
(g)-(i). However, in contrast to the behavior in the unitarity limit,
the upper peak becomes slightly broad as increasing the
temperature. This is clearly seen at the edge of the trap ($r=R_F$) in
Figs. \ref{swspat_asi10}(d)-(f).
The upper peak becomes broad as the upward dispersion of the lower
branch becomes remarkable, so that this broadening would be also caused by thermally
excited fermions.
Since the effective temperature increases as one moves away from the
trap center, this effect is most remarkable at the edge of the trap.
\par
When we define another pseudogap temperature $T^{**}$ as the temperature at which the double peak structure in LSW completely disappears, it does not coincide with the pseudogap temperature $T^*$ determined from LDOS, as shown in Fig. \ref{fig5}. As in the homogeneous case\cite{Tsuchiya}, while $T^*>T^{**}$ in the BCS side, $T^{**}$ becomes higher than $T^*$ when $(k_{\rm F}a_s)^{-1}\gesim-0.1$. Since the pseudogap is a {\it crossover} phenomenon, the pseudogap temperature may depend on what we measure. 
\par
In Fig. \ref{fig5}, we also plot $2|\mu(T_{\rm c})|$ in the BEC regime where $\mu(T_{\rm c})<0$. Since $2|\mu|$ in the BEC limit equals the binding energy of a  two-body bound state, this line physically gives the characteristic temperature where two-body bound molecules starts appearing, overwhelming thermal dissociation. Thus, the right side of $T=2|\mu(T_{\rm c})|$ in Fig. \ref{fig5} may be regarded as a gas of two-body bound molecules, rather than a Fermi atom gas. Including this, we identify the pseudogap regime as the region surrounded by $T_{\rm c}$, $2|\mu(T_{\rm c})|$, and the pseudogap temperature $T^*$ or $T^{**}$ (which depends on which we measure, LDOS or LSW).
\par
Although the phase diagram in Fig. \ref{fig5} is very similar to that for a homogeneous Fermi gas\cite{Tsuchiya}, we note that the pseudogap regime in the trapped case is narrower than that in the homogeneous case, in the sense that the ratios $T^*/T_{\rm c}$ and $T^{**}/T_{\rm c}$ in the former case are smaller than those in the latter. This is because, while pairing fluctuations are strong everywhere near $T_{\rm c}$ in the uniform case, strong pairing fluctuations are restricted to the spatial region around the trap center in the trapped case. As a result, the pseudogap phenomenon is somehow weakened by the outer region of the gas cloud.
\par
We find in Fig. \ref{fig5} that the pseudogap region is very narrow in the BCS side ($(k_{\rm F}a_s)^{-1}\lesssim -0.1$), indicating difficulty of observing the pseudogap in this weak-coupling regime. On the other hand, in the crossover region ($(k_{\rm F}a_s)^{-1}\gesim -0.1$), although the pseudogap temperature $T^*$ is still not so high, $T^{**}$ is found to give a large pseudogap regime. Since the photoemission spectrum $I_{\rm ave}({\bm q},\Omega)$ in Eq. (\ref{eq.photo}) is close to the spectrum weight $A({\bm p},\omega,r)$, the large pseudogap region in the BEC side makes us expect that the photoemission-type experiment would be useful in observing the pseudogap in trapped Fermi gases. In Sec. IV, we will check this expectation by explicitly calculating the photoemission spectrum in the BEC side.
\par
\begin{figure}
\centerline{\includegraphics[width=15cm]{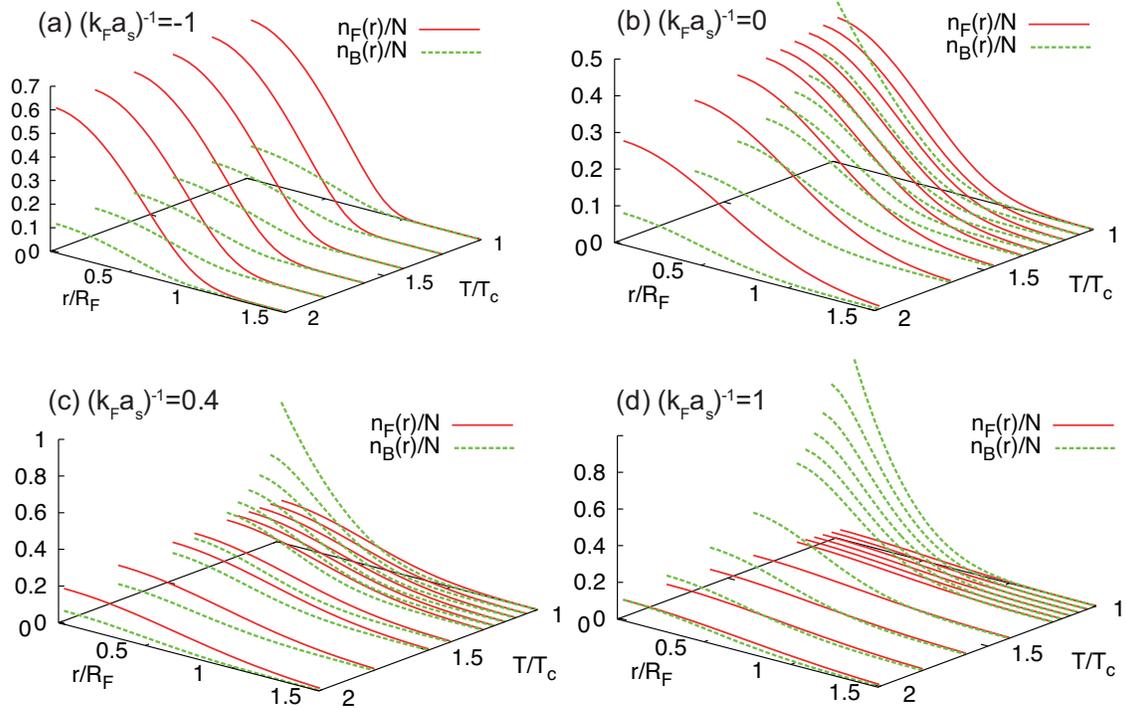}}
\caption{(Color online) Density profile of free fermion component $n_{\rm F}(r)$ and fluctuation correction $n_{\rm B}(r)$ in the BCS-BEC crossover.}
\label{fig6}
\end{figure}
\par
To see inhomogeneous pairing fluctuations in a simple manner, it is convenient to divide the atomic density profile $n(r)$ into the sum of the free fermion part $n_{\rm F}(r)=2T\sum_{\bm p,\omega_n}G^0_{\bm p}(i\omega_n,r)e^{i\omega_n\delta}$ and the fluctuation correction $2n_{\rm B}(r)$, where 
\begin{eqnarray}
n_{\rm B}(r)&=&T\sum_{\bm p,\omega_n}\left[G_{\bm p}(i\omega_n,r)-G_{\bm p}^0(i\omega_n,r)\right]e^{i\omega_n\delta}.
\label{pairdensity}
\end{eqnarray}
In the strong-coupling BEC regime where $\mu\ll-\varepsilon_{\rm F}$, $n_{\rm B}(r)$ reduces to the number of tightly bound molecules\cite{NSR}. As shown in Fig. \ref{fig6}, the fluctuation contribution $n_{\rm B}(r)$ always takes a maximum value in the trap center, as expected. While the density profile $n(r)$ is always dominated by $n_{\rm F}(r)$ above $T_{\rm c}$ in the BCS regime (panel (a)), $n_{\rm B}(r)$ in the BCS-BEC crossover region is remarkably enhanced around the trap center near $T_{\rm c}$, as shown in panels (b)-(d). 
\par
When $(k_{\rm F}a_s)^{-1}=1$ shown in panel (d), the density profile is dominated by $n_{\rm B}(r)$. In addition, we also see a cusp in $n_{\rm B}(r)$ around $r=0$ at $T_{\rm c}$. Since this cusp structure is characteristic of the LDA density profile of a Bose gas at $T_{\rm c}$\cite{Pethick}, we may regard the superfluid phase transition in this regime as a BEC of molecular bosons described by $n_{\rm B}(r)$. 
\par
\begin{figure}
\centerline{\includegraphics[width=12cm]{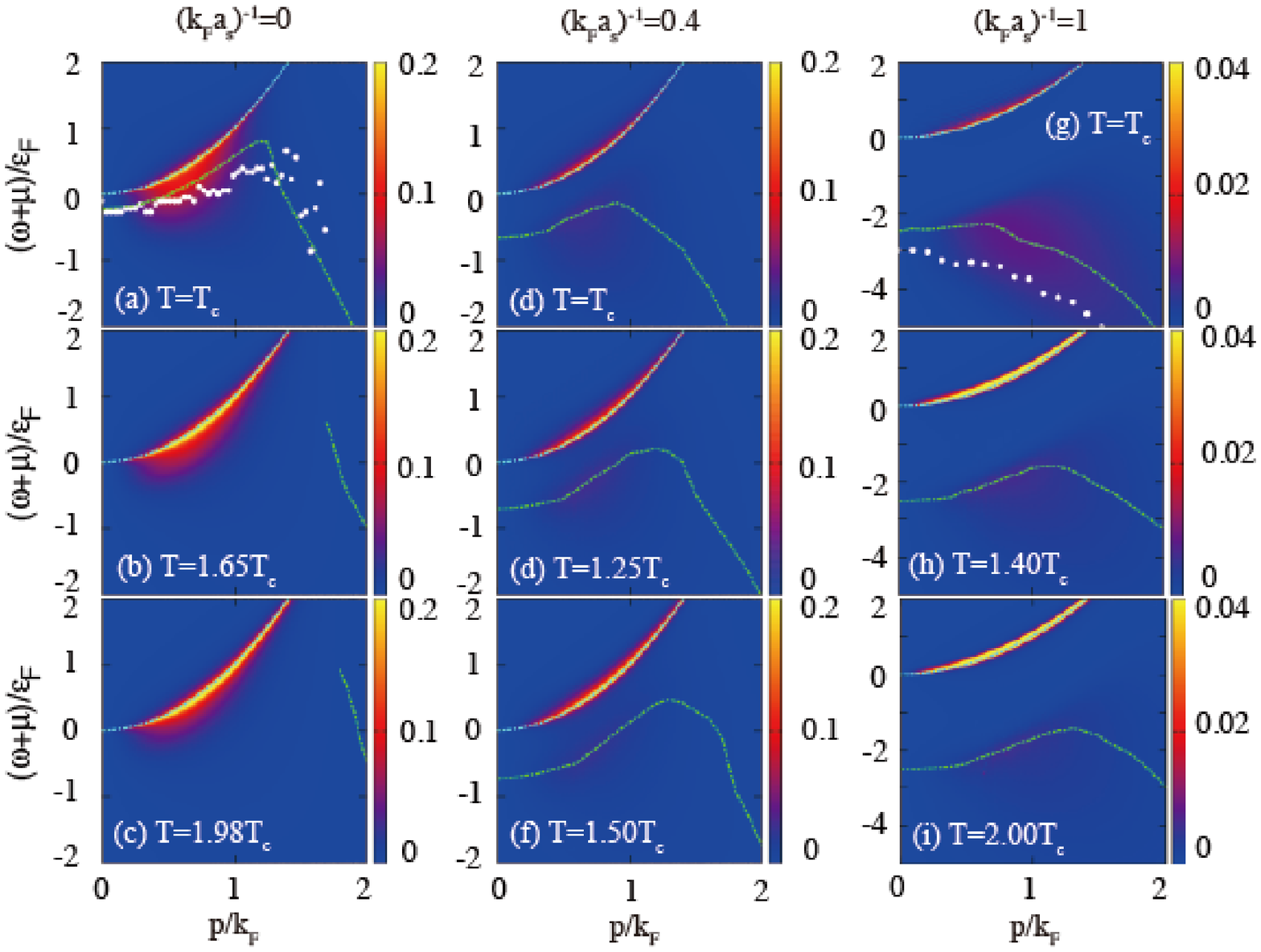}}
\caption{(Color online) Calculated intensity of photoemission spectrum $p^2\overline{A(\bm p,\omega)f(\omega)}$ above $T_{\rm c}$. The intensity is normalized by $2\pi t_{\rm F}^2/(2m)$. The spectrum has a sharp peak along the free particle dispersion $\omega=\xi_{\bm p}$ (upper dashed line). In addition, it also has a lower peak line, which is shown as the lower dashed line. Since the intensity of the lower peak is much weaker than the upper one, in some panels, one cannot see the lower one in the intensity plot. Solid circles represent the experimental data observed in Ref. \cite{Stewart}.
}
\label{fig7}
\end{figure}
\begin{figure}
\centerline{\includegraphics[width=12cm]{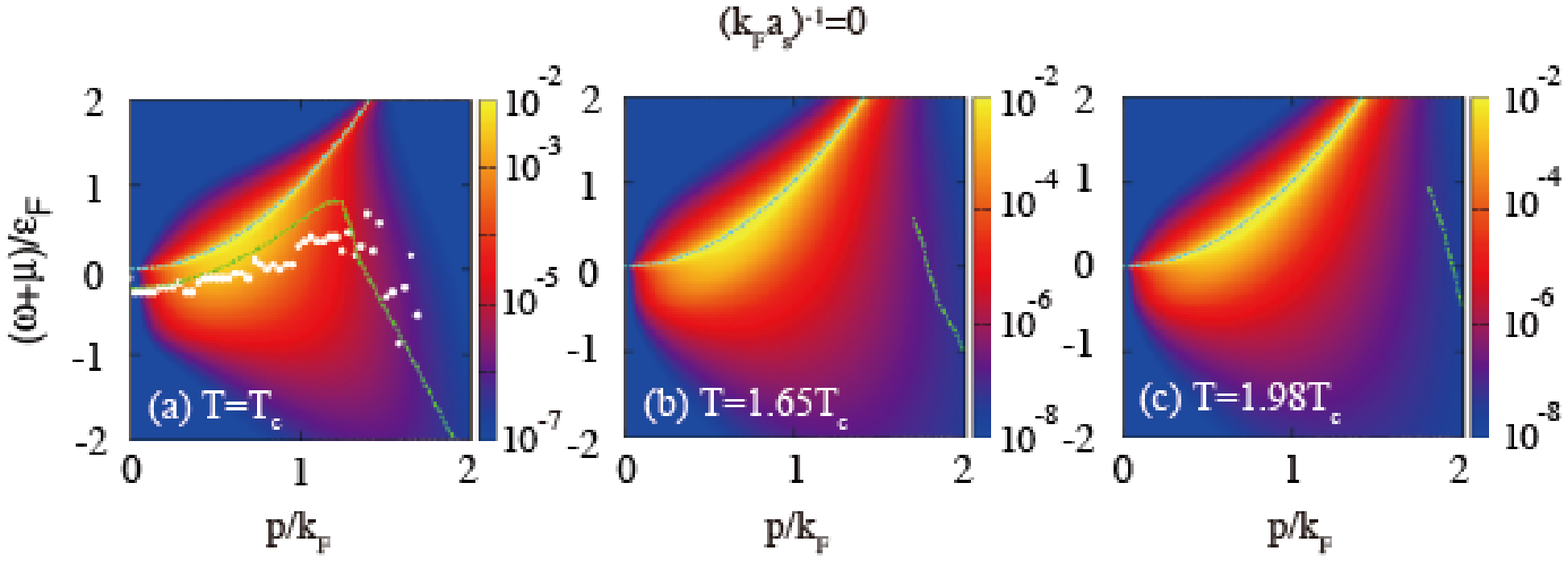}}
\caption{(Color online) Same plots as in Figs. \ref{fig7}(a)-(c), in the logarithmic scale.
 in Ref. \cite{Stewart}.}
\label{fig7_sub}
\end{figure}
\par
\section{Photoemission spectrum and pseudogap effects}
\label{section5}
Figure \ref{fig7} shows the photoemission spectrum above $T_{\rm c}$. (We also show in Fig. \ref{fig7_sub} the same plots as in Figs. \ref{fig7}(a)-(c) in the logarithmic scale, in order to clearly show the peak structure.) Although we do not explicitly show results in the weak-coupling BCS side ($(k_{\rm F}a_s)^{-1}\lesssim 0$), as expected from the phase diagram in Fig. \ref{fig5}, the photoemission spectrum in this regime is essentially the same as that for a free Fermi gas. That is, one only sees a sharp peak line along the free particle dispersion $\omega+\mu=p^2/(2m)$. On the other hand, in the crossover and BEC regime shown in Fig. \ref{fig7}, the photoemission spectrum at $T_{\rm c}$ has an upper sharp peak at $\omega+\mu\simeq p^2/(2m)$ (upper dashed line) and a lower broad peak (lower dashed line). As discussed in Ref. \cite{Tsuchiya2}, this double peak structure originates from the inhomogeneous pseudogap phenomenon seen in LSW. That is, the upper sharp peak dominantly arises from the LSW around the edge of the gas cloud where pairing fluctuations are weak, so that the upper peak position is close to the free particle dispersion. On the other hand, the lower broad peak directly reflects the lower branch in the pseudogapped LSW in the trap center where pairing fluctuations become strong near $T_{\rm c}$. 
\par
As discussed in Sec. III, the appearance of the lower branch in LSW is a direct consequence of the particle-hole coupling induced by pairing fluctuations. Because of the similarity between this coupling effect and that induced by the superfluid order parameter below $T_{\rm c}$, the lower peak line seen in Fig. \ref{fig7}(a) exhibits a back-bending behavior, being similar to the BCS hole excitation spectrum $-\sqrt{\xi_{\bm p}^2+\Delta^2}$ (although the momentum at which the back-bending occurs is different from $p=\sqrt{2m\mu}\le k_{\rm F}$ in the present case). In the BCS state, the back-bending behavior of the hole excitations spectrum $-\sqrt{\xi_{\bm p}^2+\Delta^2}$ vanishes in the strong-coupling BEC regime where $\mu<0$. This tendency can be also seen in the photoemission spectrum at $T_{\rm c}$, as shown in Figs. \ref{fig7}(a), (d), and (g). In Figs. \ref{fig7}(a) and (g), we find that the lower peak lines evaluated in the unitarity limit and BEC regime agree well with the recent experiments on $^{40}$K Fermi gases\cite{Stewart}. This indicates that the observed back-bending behavior may be understood as a signature of the pseudogap effect. Furthermore, these agreements demonstrate the validity of our strong-coupling theory for the pseudogap physics in cold Fermi gases.
\par
Since the pseudogap temperature $T^{**}$ is close to $T_{\rm c}$ in the
unitarity limit (See Fig. \ref{fig5}.), the lower spectral peak line in
the low momentum region $p/k_{\rm F}\lesssim 1.5$ soon vanishes with
increasing the temperature, as shown in Figs. \ref{fig7}(b) and
(c). However, even at $T=1.98T_{\rm c}$, panel (c) shows that the lower
peak line still remains in the high momentum region $p/k_{\rm F}\gesim
1.5$, although the peak is very broad and the intensity of the peak is
weak, as shown in Figs. \ref{fig7_sub}(b) and (c). Since this temperature
is much higher than $T^{**}=1.26T_{\rm c}$, this lower peak line
remaining at high momenta would be nothing to do with the pseudogap
effect. Instead, as pointed out in Ref. \cite{Tan,Schneider,Stewart2},
this lower spectral peak is considered to arise from the universal
behavior of a Fermi gas with a contact interaction. In this regard, we
recall that, below $T^{**}$, the back-bending behavior of the lower peak
is due to the pseudogap effect, originating from the double-peak
structure of the LSW in the trap center. On the other hand, such a
double-peak structure of the LSW is absent above $T^{**}$, so that the
back-bending behavior in the high temperature regime is purely due to
the contact interaction. That is, the origin of the back-bending
behavior of the lower peak line changes around the pseudogap temperature
$T^{**}$.
\par
When $(k_{\rm F}a_s)^{-1}=0.4$ shown in Figs. \ref{fig7}(d)-(f), the
lower peak line with the back-bending behavior can be clearly seen even at $=1.5T_{\rm c}$, because of the wide pseudogap regime in this case. (See Fig. \ref{fig5}.) As one increases the temperature, the bending position moves to higher momenta, which is consistent with the recent experiment above $T_{\rm c}$\cite{Gaebler}. 
\par
We briefly note that, although the region at $(k_{\rm F}a_s)^{-1}=1$ may be regarded as a molecular Bose gas in our phase diagram in Fig. \ref{fig5}, we still see the lower branch in Figs. \ref{fig7}(g)-(i). This indicates that many-body effects still contribute to the pair-formation to some extent even when $(k_{\rm F}a_s)^{-1}=1$. As discussed in Sec. III, the upward curve of the lower peak line in the LSW reflects the hole-like character of this branch. This behavior is also seen in Figs. \ref{fig7}(g)-(i), where the lower downward peak line around $k\simeq 0$ gradually becomes an upward one with increasing the temperature.
\par
\begin{figure}
\centerline{\includegraphics[width=10cm]{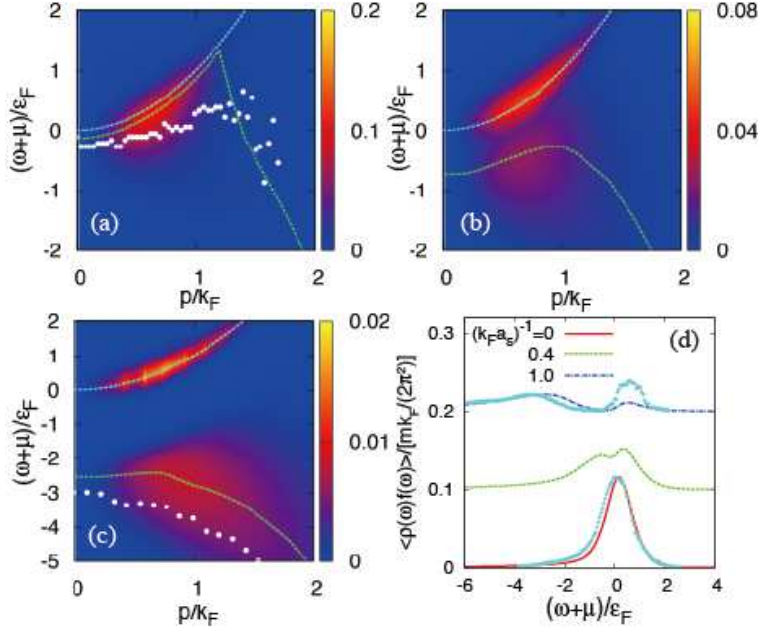}}
\caption{(Color online) Calculated photoemission spectrum $p^2\langle
 A(\bm p,\omega)f(\omega)\rangle$ affected by finite resolution, given
 by Eq. (\ref{photoave}). We take $T=T_{\rm c}$, and
 $E_r=0.2E_F$\cite{Stewart}. (a) $(k_{\rm F}a_s)^{-1}=0$. (b) $(k_{\rm
 F}a_s)^{-1}=0.4$. (c) $(k_{\rm F}a_s)^{-1}=1$. Panel (d) shows the
 effect of finite resolution on the occupied density of state, given by
 $\langle\rho(\omega)f(\omega)\rangle\equiv \sum_{\bm p}\langle A({\bm
 p},\omega)f(\omega)\rangle/(2\pi t_{\rm F}^2)$. In this figure,
 experimental data\cite{Stewart} are shown as solid circles. In panel (d), we have offset the curves by 0.1.}
\label{fig8}
\end{figure}
\par
Since the observed spectra are always affected by a finite energy resolution\cite{Stewart,Gaebler}, it is interesting to see how our results are modified by this smearing effect. To briefly examine this, we consider
\begin{eqnarray}
\langle A({\bm p},\omega)f(\omega)\rangle&=&\frac{1}{\sqrt{2\pi}E_r}\int_{-\infty}^{\infty}dz\
 \overline{A(\bm p,z)f(z)}e^{-(z-\omega)^2/(2E_r^2)},\\
\langle\rho(\omega)f(\omega)\rangle&=&\sum_{\bm p}\langle A({\bm p},\omega)f(\omega)\rangle,
\label{photoave}
\end{eqnarray}
where $E_r$ describes a finite energy resolution. We then find in Fig. \ref{fig8} that, although the resulting spectral peaks are broadened by the finite resolution, the overall spectral structure remain unchanged. Thus, the finite energy resolution ($E_r=0.2\varepsilon_{\rm F}$) is not a serious problem for the observation of the pseudogap phenomenon in the BCS-BEC regime of a cold Fermi gas.
\par
\begin{figure}
\centerline{\includegraphics[width=12cm]{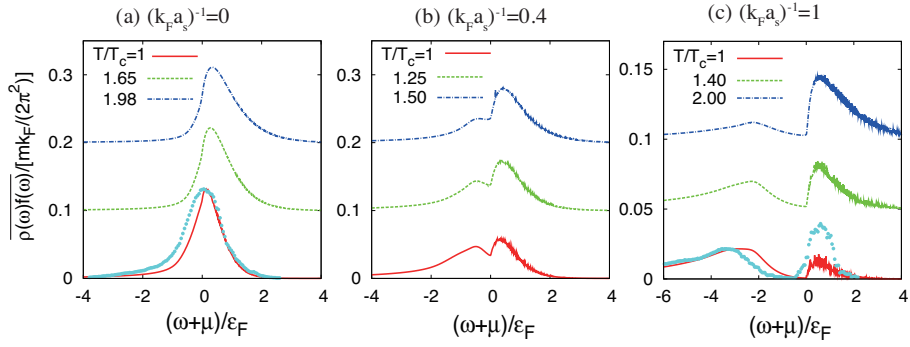}}
\caption{(Color online) Calculated occupied density of states $\overline{\rho(\omega)f(\omega)}$ above $T_{\rm c}$. In panels (a) and (c), solid circles are experimental data in Ref. \cite{Stewart}. The magnitudes of the experimental data are normalized so that their peak heights can coincide with the peak heights of our results at $T_{\rm c}$. We have offset the curves by 0.1 in panels (a) and (b), and by 0.05 in panel (c).} 
\label{fig9}
\end{figure}

Figure \ref{fig9} shows the occupied density of states $\overline{\rho(\omega)f(\omega)}$ in Eq. (\ref{eq.dos}). In the unitarity limit (panel (a)), a broad peak only appears above $T_{\rm c}$, so that we cannot see a signature of the pseudogap. However, in the BEC side, panel (b) clearly exhibits a double peak structure above $T_{\rm c}$, reflecting the pseudogap in single-particle excitations. This structure becomes more remarkable in panel (c).
\par
We compare the calculated occupied density of states with the experiment results on a $^{40}$K Fermi gas\cite{Stewart}. In the unitarity limit, the single peak structure in the unitarity limit well agrees with the observed occupied density of states, as shown in Fig. \ref{fig9}(a). In the BEC regime (panel (c)), although the relative peak height between the upper and lower peak in our result is opposite to the experimental result, their peak positions are in good agreement with the experiment. These agreements still hold, even when one takes into account the experimental finite energy resolution, as shown in Fig. \ref{fig8}(d). Thus, the present $T$-matrix theory is found to well describe strong-coupling effects on the occupied density of states.
\par

\section{Summary}
\par
To summarize, we have discussed the pseudogap phenomenon and effects of a harmonic trap in the BCS-BEC crossover regime of an ultracold Fermi gas. Extending our previous work at $T_{\rm c}$ to the region above $T_{\rm c}$, we have calculated single-particle local density of states (LDOS), as well as the single-particle local spectral weight (LSW), within the framework of the combined $T$-matrix theory with the local density approximation. We clarified how the pseudogap structures in these quantities gradually disappear with increasing the temperature. From these, we introduced two pseudogap temperatures $T^*$ (for the density of states) and $T^{**}$ (for the spectral weight).
\par
As in the case of a uniform Fermi gas, $T^*$ does not coincide with $T^{**}$. While $T^*>T^{**}$ is obtained in the weak-coupling BCS side, one finds $T^*<T^{**}$ in the strong-coupling BEC side. This means that the pseudogap temperature depends on what we measure.
\par
We have also examined the photoemission spectrum in the pseudogap regime. At present, the photoemission-type experiment has no spatial resolution, so that the observed photoemission spectra are spatially averaged ones. Although the spatial average smears inhomogeneous pseudogap effects to some extent, we showed that the double peak structure, which is characteristic of the pseudogap effect, still survives even when such a smearing effect is taken into account. The photoemission spectra, as well as the occupied density of states, calculated in the crossover and BEC regime agree well with the recent experiments on $^{40}$K Fermi gases, so that the combined $T$-matrix theory with LDA used in this paper is found to be a powerful theory to study the pseudogap physics in cold Fermi gases. Since the pseudogap temperature, which we have determined in this paper, is a key quantity in the pseudogap physics, the experimental determination of this characteristic temperature would be useful for the further understanding of strong-pairing fluctuations existing in the BCS-BEC crossover regime of a cold Fermi gases.
\par
\acknowledgements
We acknowledge D. S. Jin and J. P. Gaebler for providing us with their
experimental data. S.T. thanks A. Griffin, A. Paramekanti,
J. H. Thywissen, and T. Nikuni for fruitful discussions. Y. O. was supported by Grant-in-Aid for Scientific research from MEXT in Japan (22540412, 23104723, 23500056).

\appendix
\section{Derivation of Eq. (\ref{eq.photo})}
\label{appendix}
\par
In this appendix, we explain the outline of the derivation of Eq. (\ref{eq.photo}). In the photoemission-type experiment\cite{Stewart}, atoms in one of the two hyperfine states ($\equiv |\uparrow\rangle$) are transferred to an unoccupied hyperfine state $|3\rangle$ ($\neq|\uparrow\rangle,|\downarrow\rangle$) by rf-pulse. Although this experimental procedure is essentially the same as the rf-tunneling current spectroscopy, one can safely ignore the so-called final state interaction in the recent photoemission-type experiment on $^{40}$K\cite{Stewart}. Noting this, we consider the model Hamiltonian ${\bar H}=H+H_3+H_T$, where $H$ is given in Eq. (\ref{ham0}), and 
\begin{equation}
H_3-\mu_3 N_3=\sum_{\bm p}(\varepsilon_{\bm p}+\omega_3-\mu_3)
b_{\bm p}^\dagger b_{\bm p}
\label{third}
\end{equation}
describes the final state $|3\rangle$. Here, $b_{\bm p}$ is an
annihilation operator of a Fermi atom in $|3\rangle$, and $\omega_3$ is
the energy difference between $|\uparrow\rangle$ and
$|3\rangle$. $\mu_3$ and $N_3$ are the chemical potential and the total
number operator of atoms in the final state $|3\rangle$,
respectively. The transition from $|\uparrow\rangle$ to $|3\rangle$ is 
described by the tunneling Hamiltonian\cite{Torma,Ohashi2,He} 
\begin{equation}
H_T=t_F\sum_{\bm k}\left(e^{-i\omega_L t}b_{\bm k+\bm q_L}^\dagger c_{\bm k\uparrow}+{\rm H.c.}\right),
\label{eq.a4}
\end{equation}
where $t_F$ is a transfer matrix element between $|\uparrow\rangle$ and $|3\rangle$. $\bm q_L$ and $\omega_L$ represent the momentum and energy of the rf-pulse, respectively. As usual, we first consider a uniform Fermi gas, and then include effects of a trap by replacing $\mu_3$ with the LDA expression $\mu_3(r)=\mu_3-V(r)$. 
\par
The photoemission spectrum is conveniently described by the rf-tunneling current $I$ from the initial state $|\uparrow\rangle$ to the final state $|3\rangle$, given by
\begin{eqnarray}
I&=&
\langle{\dot N_3}\rangle=i\langle[H+H_3+H_T,N_3]\rangle=\langle \hat J\rangle,
\nonumber
\\
\hat J&=&-it_F\sum_{\bm k}
\left[
e^{-i\omega_Lt}b_{\bm k+\bm q_L}^\dagger c_{\bm k\uparrow}-{\rm H.c.}
\right].
\label{eq.a3}
\end{eqnarray}
Here, $N_3$ is the total number operator of Fermi atoms in the final state $|3\rangle$. Within the linear response theory\cite{Mahan} in terms of the tunneling Hamiltonian $H_T$ in Eq. (\ref{eq.a4}), Eq. (\ref{eq.a3}) reduces to
\begin{equation}
I=-i\int_{-\infty}^{t}dt^\prime~
\langle[\hat J(t),H_T(t^\prime)]\rangle e^{\delta t^\prime},
\label{eq.I}
\end{equation}
where $H_T(t)=e^{i(H+H_3) t}H_T e^{-i(H+H_3) t}$, and $\hat J(t)=e^{i(H+H_3) t}\hat Je^{-i(H+H_3) t}$.
\par
Equation (\ref{eq.I}) can be conveniently evaluated from the corresponding thermal Green's function by analytic continuation. That is, when one introduces the correlation function in the Matsubara formalism, 
\begin{equation}
\Lambda(i\nu_n)=\frac{t_F^2}{\beta}\sum_{\bm k,\omega_n}G_{\bm k\uparrow}(i\omega_n)G_{{\bm k}+{\bm q}_L,3}(i\omega_n+i\nu_n),
\label{eq.Lambda}
\end{equation}
(where $G_{\bm p,3}(i\omega_n)=1/[i\omega_n-(\varepsilon_{\bm p}-\mu_3)]$ is the single-particle Green's function for the final state $|3\rangle$), Eq. (\ref{eq.I}) is given by
\begin{equation}
I=-2\ {\rm Im}[\Lambda(i\nu_n\to \Omega+\mu-\mu_3+i\delta)].
\end{equation}
Here, $\Omega\equiv\omega_L-\omega_3$ is the rf-detuning. Carrying out the summation with respect to the Matsubara frequencies in Eq. (\ref{eq.Lambda}), one finds
\begin{equation}
I(\bm p,\Omega)=2\pi t_F^2 A(\bm
 p,\xi_{\bm p}-\Omega)f(\xi_{\bm p}-\Omega). 
\label{rfcurrent}
\end{equation}
In obtaining Eq. (\ref{rfcurrent}), we have assumed that (1) the photon momentum is negligibly small ($\bm q_L=0$), and (2) the final state $|3\rangle$ is initially empty ($f(\varepsilon_{\bm p}-\mu_3)= 0$). We find from Eq. (\ref{rfcurrent}) that the photoemission spectrum is related to the {\it occupied} spectral weight as 
\begin{equation}
I(\bm p,\Omega\to\xi_{\bm p}-\omega)=2\pi t_F^2A(\bm p,\omega)f(\omega).
\label{eq.emission}
\end{equation}
At $T=0$, since the Fermi distribution function $f(\omega)$ reduces to
the step function, Eq. (\ref{eq.emission}) gives the ordinary spectral
weight $A(\bm p,\omega)$ below $\omega=0$. At finite temperatures, on
the other hand, thermally excited quasiparticles contribute to the spectrum, so that the photoemission spectrum $I(\bm p,\Omega\to\xi_{\bm p}-\omega)$ in Eq. (\ref{eq.emission}) has finite intensity even for $\omega>0$.
\par
The extension of Eq. (\ref{rfcurrent}) to a trapped gas is achieved by replacing $\mu$ with the LDA expression $\mu(r)=\mu-V(r)$, leading to the {\it local} photoemission spectrum, given by
\begin{equation}
I(\bm p,\Omega,r)=2\pi t_F^2A(\bm p,\xi_{\bm p}(r)-\Omega,r)f(\xi_{\bm p}(r)-\Omega),
\label{rfLDA}
\end{equation}
where $\xi_{\bm p}(r)=\varepsilon_{\bm p}-\mu(r)$. In the current experiment\cite{Stewart}, since the rf-pulse is always applied to the whole gas cloud, the observed spectrum involves contributions from all the spatial regions of the gas cloud. To include this, we take the spatial average of Eq.(\ref{rfLDA}) as
\begin{equation}
I_{\rm ave}(\bm p,\Omega)=\frac{2\pi t_F^2}{V}\int d{\bm r}
A(\bm p,\xi_{\bm p}(r)-\Omega,r)f(\xi_{\bm p}(r)-\Omega).
\label{eq.ave}
\end{equation}
Here, $V=4\pi R_F^3/3$ is a characteristic volume of the gas cloud, where $R_F=\sqrt{2\mu/(m\omega_{\rm tr}^2)}$ is the Thomas-Fermi radius\cite{Pethick}. We emphasize that Eq. (\ref{eq.ave}) gives a proper definition of the spatially averaged photoemission spectrum, being comparable to the observed spectra by JILA group.
\par



\begin{thebibliography}{}
\bibitem{notezz} The self-energy correction in Eq. (\ref{eq.self}) also has a temperature dependence which is independent of the temperature dependence of $\mu$. However, this temperature dependence is actually weak, so that this effect is not clearly seen in Figs. \ref{fig4} and \ref{swspat_asi0}.
\end{thebibliography}
\end{document}